\documentclass{jpp}
\usepackage{subeqn}
\usepackage{epsfig}

\let\realverbatim\verbatim
\let\realendverbatim\endverbatim
\renewenvironment{verbatim}
  {\par\addvspace{6pt plus 1pt minus 2pt}\realverbatim}
  {\realendverbatim\addvspace{6pt plus 1pt minus 2pt}}

\ifprodtf \else
  \checkfont{eurm10}
  \iffontfound
    \IfFileExists{upmath.sty}
      {\typeout{^^JFound AMS Euler Roman fonts on the system,
                   using the 'upmath' package.^^J}%
       \usepackage{upmath}}
      {\typeout{^^JFound AMS Euler Roman fonts on the system, but you
                   don't seem to have the}%
       \typeout{'upmath' package installed. JPP.cls can take advantage
                 of these fonts,^^Jif you use 'upmath' package.^^J}%
       \providecommand\umu{\umu}%
      }
  \else
    \providecommand\umu{\mu}%
  \fi
\fi

\ifprodtf \else
  \checkfont{msam10}
  \iffontfound
    \IfFileExists{amssymb.sty}
      {\typeout{^^JFound AMS Symbol fonts on the system, using the
                'amssymb' package.^^J}%
       \usepackage{amssymb}%

      }{}
  \else
  \fi
\fi

\ifprodtf \else
  \IfFileExists{amsbsy.sty}
    {\typeout{^^JFound the 'amsbsy' package on the system, using it.^^J}%
     \usepackage{amsbsy}}
    {}
\fi

\newcommand{\Fig}[1]{Fig. (\ref{#1})}

\newcommand{\be}{\begin{equation}}
\newcommand{\ee}{\end{equation}}

\newcommand{\eqa}{\begin{eqnarray*}}
\newcommand{\eqe}{\end{eqnarray*}}
\newcommand{\eqnu}{\begin{eqnarray}}
\newcommand{\eqne}{\end{eqnarray}}

\newcommand{\eeq}{\end{eqnarray}}

\newdefinition{definition}[theorem]{Definition}

\title[Journal of Plasma Physics]
{Turbulent Spectra in the Solar Wind Plasma}

\author[Shaikh \& Zank]
{D\ls A\ls S\ls T\ls G\ls E\ls E\ls R \ns S\ls H\ls A\ls I\ls K\ls
H\ls$^\ast$  and G.\ls P.\ls Z\ls A\ls N\ls K\ls}
\affiliation{Department of Physics and \\
Center for Space Physics and Aeronomic Research (CSPAR),\\
University of Alabama at Huntsville, Huntsville, AL 35805. USA.\\
{$^\ast$\tt Email:dastgeer.shaikh@uah.edu}}
\date{April 21 2009; Revised on June 28, 2009; Accepted on July 6, 2009}
\pubyear{2009} % only needed when year of publication is not current year.
\volume{00}
\part{0}
\pagerange{\pageref{firstpage}--\pageref{lastpage}}
\doi{S0963548301004989}

\begin{document}

\label{firstpage}
\maketitle

\begin{abstract}

Observations of interstellar scintillations at radio wavelengths
reveal a Kolmogorov-like scaling of the electron density spectrum with
a spectral slope of $-5/3$ over six decades in wavenumber space.  A
similar turbulent density spectrum in the solar wind plasma has been
reported.  The energy transfer process in the magnetized solar wind
plasma over such extended length-scales remains an unresolved paradox
of modern turbulence theories raising the especially intriguing
question of how a compressible magnetized solar wind exhibits
a turbulent spectrum that is a characteristic of an incompressible
hydrodynamic fluid.  To address these questions, we have undertaken
three-dimensional time dependent numerical simulations of a
compressible magnetohydrodynamic fluid describing super-Alfv\'enic,
supersonic and strongly magnetized plasma. It is shown that the
observed Kolmogorov-like ($-5/3$) spectrum can develop in the solar
wind plasma by supersonic plasma motions that dissipate into highly
subsonic motion that passively convect density fluctuations.

\end{abstract}

%\tableofcontents

\section{Introduction}

It is a remarkable that observations of electron density fluctuations
in both the interstellar medium (ISM) and solar wind plasma exhibit an
omnidirectional Kolmogorov-like power spectrum $k^{-5/3}$ (Armstrong
et al 1981; Armstrong et al 1990).  There is a zoo of observations
that suggest a $k^{-5/3}$ Kolmogorov-like power spectrum of turbulence
in the solar wind plasma.  For instance, Goldstein et al (1995)
reported 0.12-second average Mariner 10 magnetometer data that exhibit
magnetic field fluctuation spectrum very close to -5/3 with a clearly
defined dissipation range at higher frequencies. The power density
spectra of magnetic field fluctuations observed by Helios 1, 2 and
Ulysses between 0.3 and 1 AU indicate the dependence of the power
spectra relative to the heliocentric distance (Bruno and Carbone
2005). The higher frequency component of the spectra is consistent
with the Kolmogorov-like 5/3 spectrum. The spectral properties of the
solar wind observations are described extensively in a comprehensive
review article by Bruno and Carbone (2005). Tu et al (1991) and
Marsch \& Tu (1990) reported a Kolmogorov-like density, proton
temperature and magnetic field fluctuations spectra in the solar wind
plasma.  Bellamy et al (2005) reported an extensive spectral analyses
of the plasma density based on the Voyager 2 spacecraft to investigate
the spectral characteristics and fluctuation level of density
turbulence from 1 to 60 AU, corresponding to the period 1977 to
1999. They find that the density spectra associated with high
frequency (above about $10^{-4}$ Hz) solar wind turbulence in the
outer heliosphere have spectral index that is close to a
Kolmogorov-like $k^{-5/3}$ power law.  While the low frequency
component of the spectrum follows a $k^{-2}$ law, the high frequency
part of the spectrum show a slight decrease in the spectral index.  A
nearly incompressible theory has been proposed to explain the observed
solar wind density spectrum by Zank \& Matthaeus (1993) and Matthaeus
et al. [1991]. The latter confirmed that density spectra near 3 AU
have a spectral index very similar to the magnetic and velocity field
spectral index. In a most detailed analysis of den- sity spectra
interior to 1 AU, Marsch and Tu [1990] showed that spectra of density
fluctuations vary as Kolmogorov-like $k^{-5/3}$ law. Leamon et al
(1998) have measured solar wind magnetic field spectrum near 1 AU
which shows a Kolmogorov-like $k^{-5/3}$ power law.

Perhaps, the most striking point about the above observations is that
the SW is a fully compressible and a magnetized medium, yet is
exhibits a Kolmogorov-like power spectrum $k^{-5/3}$ over an extended
wavenumber space. Such a spectrum is characteristic of incompressible
isotropic (wavenumbers are identical $|k|=|k_x|=|k_y|=|k_z|$) and
homogeneous (does not vary with respect to the background flow)
hydrodynamic turbulence.  The observation yields two paradoxes; (1)
why does a compressible SW fluid behave as though it were
incompressible (Goldstein et al 1995; Bruno and Carbone 2005; Tu et al
1991; Marsch \& Tu 1990; Bellamy et al 2005; Leamon et al 1998), and
(2) Why do the density fluctuations, an apparently quintessential
compressive characteristic of magnetized SW turbulence, yield a
Kolmogorov power law spectrum characteristic of incompressible
hydrodynamic turbulence?  These questions have to be answered if we
are to address the outstanding question regarding the origin of the SW
density power law spectrum.  In this article we address these issues
within the context of fully compressible Magnetohydrodynamics (MHD)
turbulence to understand how and why a supersonic, super Alfv\'enic,
and low plasma $\beta$ ($\beta$ the ratio of plasma pressure and
magnetic pressure) SW fluid should exhibit a Kolmogorov-like
wavenumber spectrum in density. Our results are valid for all
length-scales that constitute isotropic and homogeneous SW fluid.

In a compressible fluid, the characteristic motions often exceed or
are comparable to the local sound speed. Accordingly, a local
fluctuating Mach number defined as $M_{s_0}=U_0/C_{s_0},
C_{s_0}^2=\gamma p_0/\rho_0$, (where $U_0, C_{s_0}$ are the
characteristic and sound speeds of a turbulent fluid, and
$p_0, \rho_0$ and $\gamma$ are pressure, density and adiabatic index
of magnetoplasma respectively) changes. It primarily expresses the
degree of compressibility of the magnetofluid.  A compressible
magnetofluid contains fast and slow magnetoacoustic modes. By
contrast, an incompressible magnetofluid corresponds to a fluid in
which such fast-scale modes are absent and it contains purely vortical
modes. The incompressible and compressible magnetoplasma fluids
therefore differ considerably from each other.  Sound waves do not
exist in an incompressible fluid, hence the turbulent Mach number
$M_{s_0} \rightarrow 0$. Turbulent Mach number in a compressible fluid
can depend on local parameters, it is therefore useful to define a
fluctuating Mach number that varies in space and time $M_s({\bf r},
t)$ (${\bf r}=x\hat{e}_x+y\hat{e}_y + z\hat{e}_z$ is a three
dimensional space vector). The fluctuating Mach number $M_s({\bf r},
t)$ is different from the fixed Mach number $M_{s_0}$. The latter can
be viewed as a consequence of the normalizing quantities that
represent large-scale flows. We use the fluctuating (small-scale) Mach
number $M_s({\bf r}, t)$ as a prime diagnostic to investigate the
dynamics of {\it multi-scale coupling} in a super-Alfv\'enic,
supersonic, and a strongly magnetized compressible MHD plasma. All the
small-scale fluctuating parameters are measured in terms of their
respective normalized quantities.

In section 2, we describe our simulation model. Section 3 deals with
the nonlinear simulation results. The mode coupling interaction
leading to a Kolmogorov-like spectrum is described in section 4,
whereas section 5 contains summary.

\section{MHD model}
The fluid model describing nonlinear turbulent processes in the
magnetofluid plasma, in the presence of a background magnetic field,
can be cast into plasma density ($\rho_p$), velocity (${\bf U}_p$),
magnetic field (${\bf B}$), pressure ($P_p$) components according to
the conservative form
\be
\label{mhd}
 \frac{\partial {\bf F}_p}{\partial t} + \nabla \cdot {\bf Q}_p={\cal Q},
\ee
where,
\[{\bf F}_p=
\left[ 
\begin{array}{c}
\rho_p  \\
\rho_p {\bf U}_p  \\
{\bf B} \\
e_p
  \end{array}
\right], 
{\bf Q}_p=
\left[ 
\begin{array}{c}
\rho_p {\bf U}_p  \\
\rho_p {\bf U}_p {\bf U}_p+ \frac{P_p}{\gamma-1}+\frac{B^2}{8\pi}-{\bf B}{\bf B} \\
{\bf U}_p{\bf B} -{\bf B}{\bf U}_p\\
e_p{\bf U}_p
-{\bf B}({\bf U}_p \cdot {\bf B})
  \end{array}
\right] \],\\
\[ {\cal Q}=
\left[ 
\begin{array}{c}
0  \\
{\bf f}_M({\bf r},t) +\mu \nabla^2 {\bf U}+\eta \nabla (\nabla\cdot {\bf U})  \\
\eta \nabla^2 {\bf B}  \\
0
  \end{array}
\right]
\] 
and
\[ e_p=\frac{1}{2}\rho_p U_p^2 + \frac{P_p}{\gamma-1}+\frac{B^2}{8\pi}.\]
Equations (1) are normalized by typical length $\ell_0$ and time $t_0
= \ell_0/V_A$ scales in our simulations such that
$\bar{\nabla}=\ell_0{\nabla}, \partial/\partial
\bar{t}=t_0\partial/\partial t, \bar{\bf U}_p={\bf U}_p/V_A,\bar{\bf
  B} ={\bf B}/V_A(4\pi \rho_0)^{1/2}, \bar{P}=P/\rho_0V_A^2,
\bar{e}_p=e_p/\rho_0V_A^2, \bar{\rho}=\rho/\rho_0$. The bars are
removed from the normalized equations
(1). $V_A=B_0/(4\pi \rho_0)^{1/2}$ is the Alfv\'en speed. Compressible
MHD model is employed by Cho \& Lazarian (2003) to study magnetic
field spectrum in magnetized plasma. In the context of interstellar
medium, Kissmann et al (2008) find that density power spectra exhibit
a departure from the classical Kolmogorov phenomenology.

The rhs in the momentum equation denotes a forcing functions (${\bf
  f}_M({\bf r},t)$) that essentially influences the plasma momentum at
the larger length scale in our simulation model.  With the help of
this function, we drive energy in the large scale eddies to sustain
the magnetized turbulent interactions. In the absence of forcing, the
turbulence continues to decay freely. 

\section{Simulation results}
Nonlinear mode coupling interaction studies in three (3D) dimensions
are performed to investigate the multi-scale evolution of
decaying/driven compressible MHD turbulence.  The characteristic
wavenumbers $k$ and frequencies $\omega$ over which MHD is valid are
defined by $k\rho_i <1,~ \omega<\omega_{c_i}<\omega_{p_i}$, $\rho_i$
is ion gyro radius, $\omega_{ci}$ and $\omega_{p_i}$ are respectively
the ion cyclotron and ion plasma frequencies. In the simulations, all
the fluctuations are initialized isotropically with random phases and
amplitudes in Fourier space and an initial shape close to $k^{-2}$
($k$ is the Fourier mode, which is normalized to the characteristic
turbulent length-scale $l_0$). No mean magnetic or velocity fields are
assumed in the initial fluctuations, but they may be generated locally
by nonlinear interactions.  This ensures that the initial fluctuations
are fairly isotropic and no anisotrpy is introduced by the initial
data.  However a background constant magnetic field is assumed along
the $z$ direction in our simulations that is consistent with the solar
wind background constant magnetic field.  Since we are interested in a
local region of the solar wind plasma, the computational domain
employs a three-dimensional periodic box of volume $\pi^3$.  Kinetic
and magnetic energies are also equi-partitioned between the initial
velocity and the magnetic fields. The latter helps treat the
transverse or shear Alfv\'en and fast/slow magnetosonic waves on an
equal footing, at least during the early phase of the simulations.
MHD turbulence evolves under the action of nonlinear interactions in
which larger eddies transfer their energy to smaller ones through a
forward cascade. During this process, MHD turbulent fluctuations are
dissipated gradually due to the finite Reynolds number, thereby
damping small scale motion as well. This results in a net decay of
turbulent sonic Mach number $M_s$ which was demonstrated in Shaikh \&
Zank (2006).  The turbulent sonic Mach number continues to decay from
a supersonic ($M_s>1$) to a subsonic ($M_s<1$) regime. This indicates
that dissipative effects predominantly cause supersonic MHD plasma
fluctuations to damp strongly leaving primarily subsonic fluctuations
in the MHD fluid.  Note that the dissipation is effective only at the
small-scales, whereas the large-scales and the inertial range
turbulent fluctuations remain unaffected by direct dissipation. Since
there is no mechanism that drives turbulence at the larger
scales. Spectral transfer in driven turbulence, in general, follows a
similar cascade process as in the decaying turbulence case, but it may
be more complex depending upon what modes (irrotational or solenoidal)
are being excited. The energy containing eddies, the large-scale
energy simply migrates towards the smaller scales by virtue of
nonlinear cascades in the inertial range and is dissipated at the
smallest turbulent length-scales.  Spectral transfer in globally
isotropic and homogeneous hydrodynamic and magnetohydrodynamic
turbulence is the widely accepted paradigm (Kolmogorov 1941;
Iroshnikov 1963; Kraichnan 1965) that leads to Kolmogorov-like energy
spectra.  The most striking effect, however, to emerge from the decay
of the turbulent sonic Mach number is that the density fluctuations
begin to scale quadratically with the subsonic turbulent Mach number
as soon as the compressive plasma enters the subsonic regime,
i.e. $ \delta \rho
\sim {\cal O}(M_s^2)$ when $M_s<1$ (Shaikh \&
Zank (2006)). This signifies an essentially {\it weak} compressibility
in the magnetoplasma, and can be referred to as a {\it nearly
incompressible} state (Matthaeus et al 1988; Zank \& Mstthaeus 1990,
1993).

\begin{figure}
\begin{center}
\includegraphics[width=9.cm]{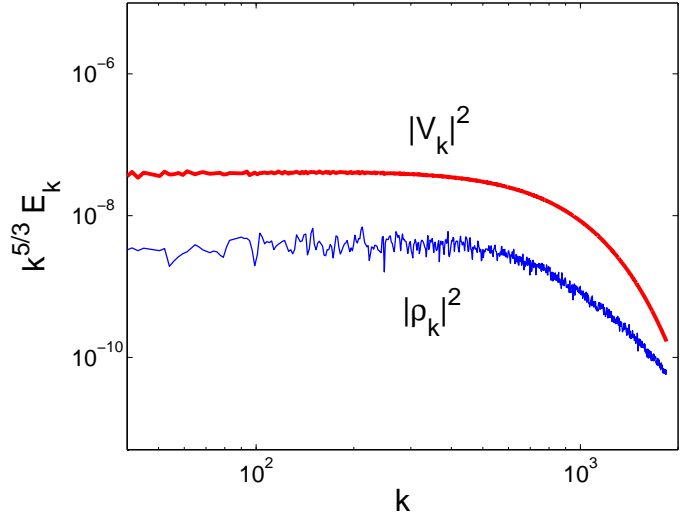}% Here is how to importEPS art
\end{center}
\caption{\label{fig2} (Top curve) Velocity fluctuations are dominated
by shear Alfv\'enic motion and thus exhibit a Kolmogorov-like
$k^{-5/3}$ spectrum, where $k$ is the Fourier mode. (Lower curve)
Density fluctuations are passively convected by the nearly
incompressible shear Alfv\'enic motion and follow a similar spectrum
in the inertial range. The numerical resolution in 3D is
$512^3$. Turbulent Reynolds numbers are
$\tilde{R}_e=\tilde{R}_m\approx 200$.}
\end{figure}

As a direct consequence of the magnetoplasma being near
incompressible, the density fluctations exhibit a weak compressibility
in the gas and are convected predominantly passively in the background
incompressible fluid flow field. This hypothesis can be verified
straightforwardly by investigating the density spectrum which should
be slaved to the incompressible velocity spectrum. This is shown in
\Fig{fig2} which illustrates that the density fluctuations follow the
velocity fluctuations in the inertial regime after the long time
(several Alfv\'en transit time) evolution of MHD turbulence. The
transition of compressible magnetoplasma from a(n) (initial)
supersonic to a subsonic or nearly incompressible regime is
gradual. This means that the magnetofluid contains supersonic, and
super Alfv\'enic modes initially in which highly compressible density
fluctuations do not follow the velocity spectrum. It is the eventual
decay of the turbulent Mach number to a subsonic regime that is
responsible for the density fluctuations following the velocity
fluctuations. In the subsonic regime, the compressibility weakens
substantially so that the density fluctuations are advected only
passively.  A passively convected fluid exhibits a similar inertial
range spectra as that of its background flow field (Macomb
1990). Likewise, subsonic density fluctuations in our simulations
exhibit a Kolmogorov-like $k^{-5/3}$ spectra similar to the background
velocity fluctuations in the inertial range.  This, we believe,
provides a plausible explanation for the Kolmogorov-like density
spectrum observed in the solar wind plasma i.e. they are convected
passively in a field of nearly incompressible velocity fluctuations
and acquire identical spectral features [as shown in \Fig{fig2}]. The
passive scalar evolution of the density fluctuations is associated
essentially with incompressiblity and can be understood directly from
the continuity equation as follows. Expressing the fluid continuity
equation as $(\partial_t + {\bf U} \cdot \nabla) \ln \rho = - \nabla
\cdot {\bf U} $, where the rhs represents compressiblity of the
velocity fluctuations, shows that the density field is advected
passively when the velocity field of the fluid is nearly
incompressible with $\nabla \cdot {\bf U} \simeq 0$. This is in part
demonstrated in Shaikh \& Zank (2007).

It is noted that the energy spectra depicted in \Fig{fig2} are
globally isotropic. It can be locally anisotropic if there exists a
self-consistently generated large-scale or mean magnetic field
i.e. $\lambda_\parallel \ne \lambda_\perp$, where $\lambda_\parallel$
and $\lambda_\perp$ are the length-scales of turbulent eddies oriented
relative to the mean magnetic field.  Local anisotropy in solar wind
turbulence can be mediated by nonlinear interactions that lead to the
formation of large-scale or mean magnetic field. These large-scale
local eddies act as guide fields for small-scale turbulent
fluctuations and excite Alfv\'en waves locally. The Alfv\'en waves can
potentially inhibit turbulent cascades along the direction of
propagation.  Consequently, energy transfer across and along the {\it
local} mean magnetic fields occurs at different rates and thus
$\lambda_\parallel > \lambda_\perp$.  In locally isotropic
magnetofluid turbulence, $\lambda_\parallel \simeq \lambda_\perp$.
Any disparity between the parallel and the perpendicular local
turbulent fluctuations (or the scale-dependent anisotropy) is revealed
quantitively from the second order structure function. The second
order structure function, corresponding to the energy associated with
the Kolmogorov-like spectrum, describes turbulent eddy structure in
the {\it local} real space, where $\lambda_\parallel$ and
$\lambda_\perp$ are respectively the parallel and perpendicular
length-scales of the eddies relative to the local mean magnetic field.
It is defined as $S_2(\ell) = \langle |{\bf U}({\bf x}+\ell, t)-{\bf
U}({\bf x}, t)|^2 \rangle$, where ${\bf U}$ are velocity or magnetic
field fluctuations, $\ell$ is either parallel or perpendicular
wave-length, and $\langle \cdots \rangle$ represents an average over
all {\bf x}.  The local anisotropy in the turbulent cascade is
demonstrated in Shaikh \& Zank (2007).  The local large-scale
fluctuations in solar wind turbulence exhibits anisotropy, whereas the
global energy spectrum [see \Fig{fig2}] is representative of
Kolmogorov-like isotropic turbulence.  It has been argued that an
anisotropic cascade in magnetofluid turbulence modifies the
Kolmogorov-like $k^{-5/3}$ spectrum to the Iroshnikov-Kriachnan
(IK)-like $k^{-3/2}$ spectrum (Kolmogorov 1941; Iroshnikov 1963;
Kraichnan 1965). The physical arguments invoked to justify the IK-like
spectrum are based on the self-interaction of colliding Alfv\'enic
wave packets that tend to flatten out the MHD spectrum (Ng et al
2003), yielding a $k^{-3/2}$ spectrum.  This issue has been a subject
of intense debate and is beyond the scope of this work. Our results
describing in \Fig{fig2} are further consistent with observational
evidence (Goldstein et al 1995; Bruno and Carbone 2005; Tu et al 1991;
Marsch \& Tu 1990; Bellamy et al 2005; Leamon et al 1998).

\begin{figure}
\begin{center}
\includegraphics[width=7.0cm,height=6.0cm]{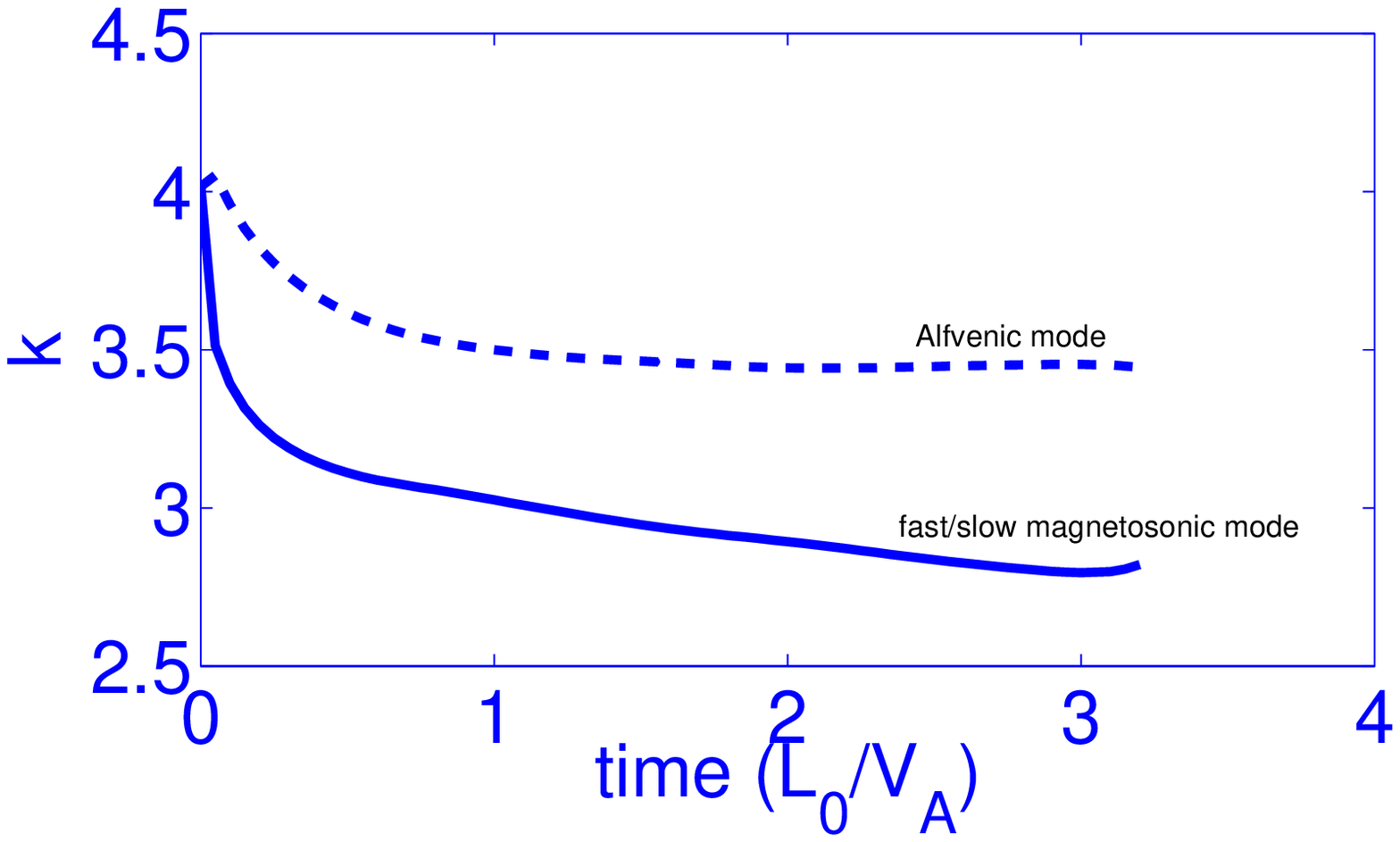}% Here is how to import EPS art
\end{center}
\caption{\label{fig3} Spectral energy transfer among shear Alfv\'en
modes ($\langle k_{\rm SAM} \rangle^2 \simeq \sum_{\rm \bf k} |i {\rm
\bf k} \times {\bf U}_{\bf k}|^2/ \sum_{\rm \bf k} |{\bf U}_{\bf k}|^2
$) and slow/fast magnetosonic waves ($\langle k_{\rm SFM} \rangle^2
\simeq \sum_{\rm \bf k} |i {\rm \bf k} \cdot {\bf U}_{\bf k}|^2/
\sum_{\rm \bf k} |{\bf U}_{\bf k}|^2 $) during a 3D dissipative
compressible MHD simulation is shown.  These modes have identical
energy initially so that ${\rm k}_{\rm SAM} \approx {\rm k}_{\rm
SFM}(t=0)$.  As time progresses, turbulence decay and spectral
transfer due to ${\rm k}_{\rm SFM}$ is suppressed significantly. }
\end{figure}

\section{Mode coupling interactions}
It is important to investigate the mode coupling interactions in order
to understand the turbulent cascade of energy that predominantly
emigrates, in our simulations, from the compressible modes to the
incompressible modes. This is because the energy cascades are
dominated by the processes that lead to the incompressible-like
inertial range spectra in \Fig{fig2}. Alternatively, we seek to
understand how an initially non-solenoidal velocity field evolves
towards a solenoidal field as it explains the transition of the
compressible magnetoplasma from a supersonic to a subsonic or nearly
incompressible state that yields a Kolmogorov-like $k^{-5/3}$ density
spectrum. Since a non-solenoidal velocity field consists of
compressible (fast/slow magnetosonic) modes, its eventual
transformation into incompressible (Alfv\'en) MHD modes therefore
requires a detailed understanding of nonlinear mode interactions and
the subsequent energy cascade processes. To understand the nonlinear
mode coupling between these MHD modes, we introduce diagnostics that
distinguish energy cascades between Alfv\'enic and slow/fast
magnetosonic fluctuations.  Since Alfv\'enic fluctuations are
transverse, the propagation wave vector is orthogonal to the
oscillations i.e. ${\bf k} \perp {\bf U}$, and the average spectral
energy contained in these (shear Alfv\'enic modes $\langle k_{\rm SAM}
\rangle$) fluctuations can be computed from the orthogonal
fluctuations.  On the other hand, slow/fast magnetosonic modes
$\langle k_{\rm SFM} \rangle$ propagate longitudinally along the
fluctuations, i.e. ${\bf k} \parallel {\bf U}$, and thus their
contribution is determined from the longitudinal oscillations.  The
evolution of the modal energy is depicted in \Fig{fig3}. Although the
modal energies in $k_{\rm SAM} $ and $k_{\rm SFM}$ modes are identical
initially, the disparity in the cascade rate causes the energy in
longitudinal fluctuations to decay far more rapidly than the energy in
the Alfv\'enic modes. The Alfv\'enic modes, after a modest initial
decay, sustain the energy cascade processes by actively transferring
spectral power amongst various Fourier modes in the stationary state.
By contrast, the fast/slow magnetosonic modes progressively weaken and
suppress the energy cascades. The $k_{\rm SFM}$ mode represents
collectively a dynamical evolution of small-scale fast plus slow
magnetosonic cascades and does not necessarily distinguish the
individual constituents (i.e. the fast and slow modes) due to their
wave vector alignment relative to the magnetic field.  The physical
implication, however, that emerges from
\Fig{fig3} is that the fast/slow magnetosonic waves {\it do not}
contribute efficiently to the energy cascade process, and that the
cascades are governed predominantly by non-dissipative Alfv\'enic
modes that survive collisional damping in compressible MHD
turbulence. This suggests that because of the decay of the fast/slow
magnetosonic modes in compressible MHD plasmas, supersonic turbulent
motions become dominated by subsonic motions and the nonlinear
interactions are sustained primarily by Alfv\'enic modes thereafter,
the latter being incompressible. The effect of inhibiting the
fast/slow magnetosonic wave cascade is that the compressible
magnetoplasma relaxes dynamically to a nearly incompressible (NI)
state in the subsonic turbulent regime, and the solenoidal component
of the fluid velocity makes a negligible contribution
i.e. $\nabla \cdot {\bf U} \ll 1$, but not $0$. The nearly
incompressible state, leading to a Kolmogorov-like $k^{-5/3}$ spectrum
of density fluctuations, described in our simulations is further
consistent with the theoretical predictions of Zank \& Matthaeus (1990,
1993).

\section{Summary}

Within the paradigm of our model, we find that the nonlinear
interaction time for Alfv\'en waves increases compared to that of the
(magneto)acoustic waves. Consequently, the plasma motion becomes
increasingly incompressible on Alfv\'enic time scales and low
plasma-$\beta$ fluctuations are eventually transformed into high
plasma-$\beta$ fluctuations.  During this gradual transformation to
incompressibility, the compressible fast/slow magnetosonic modes do
not couple well with the Alfv\'en modes.  The cascades are therefore
progressively dominated by shear Alfv\'en modes, while the
compressible fast/slow magnetosonic waves suppress nonlinear cascades
by dissipating the longitudinal fluctuations. This physical picture
suggests that a nearly incompressible state develops naturally from a
compressive SW magnetoplasma and that the density fluctuations,
scaling quadratically with the subsonic turbulent Mach number, exhibit
a characteristic Kolmogorov-like $k^{-5/3}$ spectrum that results from
passive convection in a field of nearly incompressible velocity
fluctuations.

 The support of NASA(NNG-05GH38) and NSF (ATM-0317509) grants is  acknowledged.

%\label{lastpage}
\end{document}